\documentclass[reprint,
]{revtex4-2}

\usepackage{graphics}%
\usepackage{dcolumn}%
\usepackage{tikz}
\usepackage{bm}%
\usepackage{amsmath}

\usepackage{hyperref}%

\graphicspath{{figures/}}

\begin{document}

\preprint{APS/123-QED}

\title{Persistent versus dissipative Peltier effect in a topological quantum thermocouple}%

\author{Marco A. Jimenez-Valencia}
  \affiliation{%
 Department of Physics, University of Arizona, 1118 E. 4 th Street, Tucson, Arizona 85721 
}%
  \email{marcojv@arizona.edu}
  \author{Yiheng Xu}
  \affiliation{%
 Department of Astronomy \& Astrophysics, University of California, San Diego, La Jolla, California 92093, USA
}%
\author{Charles A. Stafford}%

\affiliation{%
 Department of Physics, University of Arizona, 1118 E. 4 th Street, Tucson, Arizona 85721 \ 
}%

\date{\today}%
                                            
\begin{abstract}
The Aharonov-Bohm (AB) effect on the thermoelectric properties of three-terminal quantum devices is investigated.
Thermodynamic relations among the linear-response coefficients of these %
devices 
are derived and interpreted.
General expressions are derived using nonequilibrium Green's functions, and applied to calculate the thermoelectric response
of a model quantum thermocouple.
It is shown that the AB effect can generate a large thermoelectric response in a device with particle-hole symmetry, which nominally has zero Seebeck and Peltier coefficients.  In addition to modifying the external electric and thermal currents of the device, the AB effect also induces persistent electric and thermal currents.   
One might expect that a persistent electric current in a quantum thermocouple, through the %
Peltier effect, could lead to persistent Peltier cooling, violating the 1st and 2nd Laws of Thermodynamics.  However, this apparent paradox is resolved by elucidating the distinction between persistent and dissipative currents in quantum thermoelectrics.

\end{abstract}

\maketitle

\section{Introduction}
Understanding the behavior of nonequilibrium quantum systems beyond the linear-response regime is of critical importance to assessing the performance and possible viability of all sorts of quantum machines.
Various authors have described novel phenomena arising %
with different builds and purposes, ranging from general statements about quantum limits on information flow as the work by Pendry \cite{pendry1983}, large thermoelectric effects in single-molecule junctions as found by Bergfield et al.\ \cite{Bergfield2010} to particular cases such as a seemingly paradoxical response presented by Whitney et al. \cite{whitneythermo1} by constructing a quantum thermocouple and deriving an exotic state of transport in the system, consequence of its non-equilibrium nature. Many others have studied three-terminal quantum systems, but mostly in the linear regime \cite{jiangPRApp2017,sanchezPE2016,Prosen}. %

The eventual realization at scale of quantum machines and the relationship between thermodynamics and information \cite{berutExperimentalVerificationLandauers2012, koskiExperimentalRealizationSzilard2014, liuPeriodicallyDrivenQuantum2021, Toyabi2010_Szilard_exp,Goold_2016, Seifert_2012} (of special significance in quantum computing) highlight the importance of having a full theoretical framework to keep close track of their internal dynamics of particles, energy and heat. In addition to this, the need of thermal management in microelectronics is projected to approach molecular lengths \cite{microheatrev}, and so more complete descriptions of thermoelectric effects in the quantum regime can be of value for the semiconductor industry as well. 

In this work, we %
show %
that by considering all currents flowing in a system far from equilibrium due to coupling with particle and thermal reservoirs, the introduction of a topological field has the two-fold consequence of modifying its thermoelectric properties and producing a current of particles and entropy previously unaccounted for.

To study these currents we model one of the most basic elements in thermoelectrics, a thermocouple, which is a three-terminal device with electric source and drain terminals and a floating electric terminal that can be either heated or cooled by the Peltier effect.  A classical thermocouple consists of two junctions in series, but a quantum thermocouple can also include a direct electrical path between source and drain, leading to a multiply connected topology which can be threaded by a magnetic flux, leading to the Aharonov-Bohm effect. A derivation and interpretation of new thermodynamic relations among the linear response coefficients of the three-terminal system, which resemble their two-terminal analog, is presented. 

An expression of the ``persistent'' particle and entropy currents in steady state for non-interacting systems in and out of equilibrium connected to multiple reservoirs is derived in the framework of the Non-Equilibrium Green's Function theory (NEGF). We further show how, if adopting the customary view of dissipative currents to the unaccounted persistent currents in such a system, we arrive at a paradox, solved by a reinterpretation on the nature of these different flows.

These distinct types of currents are contrasted in their magnitude and behavior under different parameters and bias conditions and found to be of similar magnitude, highlighting their importance in the description of local flows within the system.

Although it has been argued that the efficiency bounds of a quantum refrigerator in a three-terminal system are independent on the magnetic field \cite{whitney2016}, in this paper we present an example of a quantum system that behaves as a thermocouple, while showcasing the repercussions of the Aharonov-Bohm effect, which can enhance its thermoelectric properties, or even induce them when there are none. We determine that in this last case the induced thermoelectric effect can be of similar magnitude as in a system with built-in n-type and p-type properties.

\section{Linear thermoelectric response}
The particle %
($\nu=0$) and heat currents ($\nu=1$) flowing into a reservoir $\alpha$ through a system connected to multiple reservoirs $\beta$ with corresponding chemical potential $\mu_\beta$ and temperature $T_\beta$ can be described in the linear-response regime in electrical and thermal bias by \cite{LarsOns},
\begin{equation}
I_\alpha^{(\nu)}=\sum_\beta\left(\mathcal{L}_{\alpha\beta}^{(\nu)}\left(\mu_\beta-\mu_\alpha\right)+\mathcal{L}_{\alpha\beta}^{(\nu+1)}\frac{T_\beta-T_\alpha}{T_0}\right), \label{line1}
\end{equation}
where $\mathcal{L}_{\alpha\beta}^{(\nu)}$ are termed Onsager linear-response coefficients, related among each other by thermodynamic laws, and $T_0$ is the equilibrium temperature of the system.  The electric current flowing into reservoir $\alpha$ is then $qI_\alpha^{(0)}$, where $q=-e$ is the electron charge.

A superficial examination of this equation shows that electric bias alone could induce heat currents, and likewise, thermal bias alone could induce electric currents. This phenomenon is called thermoelectricity and has been observed and studied since the 19th century \cite{seebeck}. 

The case of interest for this work is the Peltier effect which consists on harnessing electric currents to produce heat currents. The ratio of heat to electric current flowing between reservoirs $\alpha$ and $\beta$, at equal temperatures, can be used to define a Peltier coefficient for the junction between those reservoirs
\begin{equation}
\Pi_{\alpha\beta}=\frac{\mathcal{L}_{\alpha\beta}^{(1)}}{q\mathcal{L}_{\alpha\beta}^{(0)}}, \label{piclas}
\end{equation}
which represents the quantity of heat carried per unit charge from one reservoir to another.

\begin{figure}[h]

\resizebox{.8\linewidth}{!}{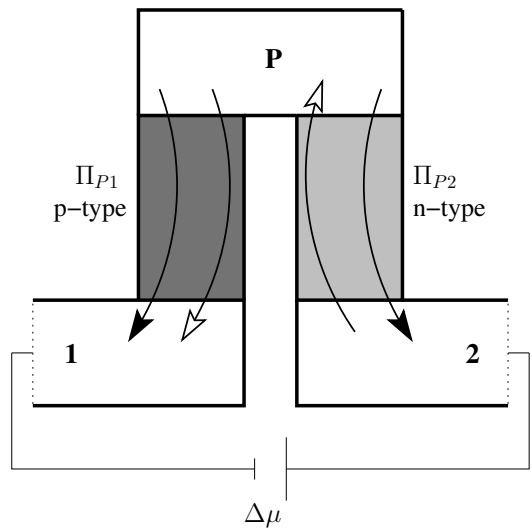}
	\caption{ \label{fig:modelclass} %
 Diagram of a conventional thermocouple with $\Pi_{P1}>0$, $\Pi_{P2}<0$. The lines with hollow arrowheads represent electric current, while the filled arrowheads, heat currents.}
\end{figure}

The Peltier effect in a three-terminal system can be achieved by setting a common temperature in a disposition such as the one shown in Fig.\ \ref{fig:modelclass}, in which $\Pi_{P1}>0$ (p-type material), $\Pi_{P2}<0$ (n-type material).

The most common system in which to observe the Peltier effect is a thermocouple which consists of a junction of a p-type and an n-type material to a common metallic contact $P$, and each of which is in turn connected individually to two other leads 1 and 2, respectively. By setting an electric bias in the circuit, the heat current from 2 to $P$ is $-\Pi_{P2}qI^{(0)}_{2}$, while from $P$ to 1 is $\Pi_{P1}qI^{(0)}_{2}$. In order to balance the heat currents and to ensure energy conservation, $(\Pi_{P2}-\Pi_{P1})qI^{(0)}_{2}$ must be extracted from reservoir $P$, as schematically depicted in Fig.\ \ref{fig:modelclass}.

In this case $\Pi_{P1}>0$ and $\Pi_{P2}<0$, and making $P$ electrically neutral  $\left( \sum_{\alpha}I^{(0)}_{P\alpha}=0 \right)$, the total heat current to $P$ is 
\begin{equation}
I_P^{(1)}=\sum_\alpha I^{(1)}_{P\alpha}=-(\Pi_{P2}-\Pi_{P1})qI^{(0)}_{2}<0.
\end{equation}
The system acts as a cooling device, extracting heat from reservoir $P$ by supplying electrical work through the electric bias between $1$ and $2$ which causes electric current to flow from contact 2 to contact 1.
Refrigerators like these have been studied since the mid 20th century, but the typical coefficients of performance of Peltier coolers are limited compared to vapor-compressing refrigerators \cite{pelthist}.

\subsection{\label{section:paradox}Persistent cooling paradox}
As early as the 1960s \cite{chemists1,chemists2}, a description of persistent currents, that is, perpetually circulating currents in a system that can be present even in equilibrium conditions, has been explored \cite{buttikerpers}.
What would happen to the heat extraction due to the Peltier effect in the presence of such a persistent current? Recall that this extraction is proportional to the current.

For instance, consider a system like the one depicted in Fig.\ \ref{fig:model2}, where the n and p-type materials have been replaced by quantum dots with resonance energies $\varepsilon_1$ and $\varepsilon_2$, where additionally a magnetic field is set and contained in the system, in the space between the leads. The topology now provides a closed path for carriers to be affected by the threading of a magnetic flux through it by the Aharonov-Bohm effect, inducing a persistent current.

 Due to the disposition of the system, without particle-hole symmetry ($\epsilon_{1}\neq\epsilon_{2}$), the system is analogous to a thermocouple.\ When $\epsilon_{2}>\mu_{0}>\epsilon_{1}$, that is, when the quantum dot at a higher resonance $\epsilon_{2}$ acts as a n-type material and the one with lower resonance energy $\epsilon_{1}$ as a p-type material, then $\Pi_{P1}>0$ and $\Pi_{P2}<0$.
 
The presence of a magnetic flux threading the system, and the fact that there is a closed circuit that carriers can traverse, breaks the time-reversal symmetry by inducing a persistent current $I_{pers}^{(0)}$ due to the Aharonov-Bohm effect \cite{AltshulerImryGeffen91}. Since this topological current can be present in the system even when there is no electric bias, and the path that encloses this magnetic field contains the reservoir $P$, can it produce a persistent heat current to cool the probe indefinitely?

If we attempt to consider this current as being of the same \textit{kind} as the dissipative current described for example by Eq.\ \eqref{line1}, then under these conditions, a paradox arises; since the closed circuit goes through the reservoir $P$, and there is no a priori reason why the Peltier coefficient difference $\Pi_{P2}-\Pi_{P1}$ cancels out, the logic of dissipative currents implies
\begin{equation}
I_{P}^{(1)}\stackrel{?}{=}(\Pi_{P2}-\Pi_{P1})qI_{pers}^{(0)}. \label{paradox}
\end{equation}
This would represent a persistent heat current that may be set to cool a reservoir indefinitely without the need of providing power into the system (other than the necessary work to initially set up the persistent current), violating the First and Second Laws of Thermodynamics.

\subsection{Thermodynamic relations of linear-response coefficients in three-terminal systems}
\begin{figure}[htbp]
\begin{center}
\resizebox{.8\linewidth}{!}{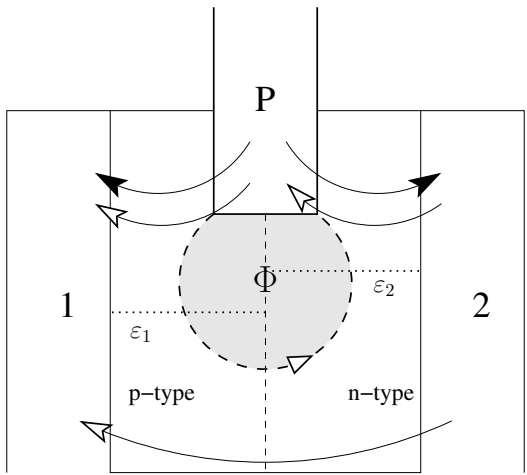}
\caption{\label{fig:model2} %
Schematic of the proposed three-terminal system to act as a thermocouple. Lines with hollow arrowheads represent particle currents, and with filled arrowheads represent heat currents. The dashed arrow represents the topological particle current due to the AB effect. Note $\Pi_{P1}>0$, $\Pi_{P2}<0$.}
\end{center}
\end{figure}

In order to explore the role of persistent currents far from equilibrium and understand why the aforementioned persistent cooling is unattainable, we must study quantum transport in three-terminal systems. The quantum transport of charge and energy can also be characterized in linear response by %
$\mathcal{L}_{\alpha\beta}^{(\nu)}$ coefficients. 

The proposed model is a quantum system connected to three reservoirs, which entails the possibility of coherent tunneling of charge carriers directly from reservoir $1$ to $2$, with corresponding temperatures $T_1$ and $T_2$, and chemical potentials $\mu_1$ and $\mu_2$, as depicted in Fig. \ref{fig:model2}. The carriers' capacity of tunneling distinguishes this scenario from two successive junctions with a common reservoir as in Fig.\ \ref{fig:modelclass}.

The Second Law of Thermodynamics states
\begin{equation}
    \sum_{\alpha} \frac{d S_\alpha}{d t} \geq 0,
    \label{eq:Sdot}
\end{equation}
where $\alpha=1,2,P$ in this three-terminal system. Eq.\ \eqref{eq:Sdot} implies the positivity of the $\mathcal{L}$ matrix, which for 
the case where terminals 1 and 2 are held at a common temperature $T_1=T_2=T_0$, 
may be expressed as
\begin{equation}
\tilde{\mathcal{L}}_{12}^{(0)}\kappa_P-q^2\left(\tilde{\mathcal{L}}_{12}^{(0)}\right)^2S_{P,12}\Pi_{P,12}  \geq 0. \label{tineq}
\end{equation}
The terms in this expression are %
linear thermoelectric coefficients for this three-terminal system %
and can be interpreted in direct correspondence with a two-terminal system.

With the definition $\mathcal{L}_{P}^{(\nu)}\equiv \sum_\alpha \mathcal{L}_{P\alpha}^{(\nu)}=\sum_ \alpha\mathcal{L}_{\alpha P}^{(\nu)}$ from conservation of probability, the term
\begin{equation}
	\kappa_P=\frac{1}{T_0}\left[\mathcal{L}_P ^{(2)}-\frac{\left(\mathcal{L}_P^{(1)}
		\right)^2}{\mathcal{L}_P^{(0)}}\right]
\end{equation}
corresponds to the parallel thermal conductance of probe $P$ to $1$ and $2$. This is the proportionality factor of the heat current into $P$ to the thermal bias $\Delta T=T_0-T_P$.  We take $P$ to be an electrically floating probe with $I_P^{(0)}=0$.
\begin{equation}
\tilde{\mathcal{L}}_{12}^{(0)}=\mathcal{L}_{12}^{(0)}+\frac{\mathcal{L}_{1P}^{(0)}\mathcal{L}_{P2}^{(0)}}{\mathcal{L}_{P}^{(0)}}, \label{effconduc}
\end{equation}
is the effective particle conductance from $1$ to $2$ \cite{PhysRevB.40.3409,5390027}, where the first term corresponds to coherent conductance and the second the incoherent conductance of carriers traveling first from $1$ to $P$ and then to $2$. This is the constant of proportionality for particle current into reservoir $1$, $I_1^{(0)}$, at electrical bias $\Delta\mu=\mu_2-\mu_1$ when $P$ %
is held at the common temperature $T_P=T_0$. %
\begin{equation}
	S_{P,12}=\frac{1}{T_0}\frac{\mathcal{L}_{1P}^{(0)}\mathcal{L}_{2P}^{(1)}-\mathcal{L}_{1P}^{(1)}\mathcal{L}_{2P}^{(0)}}{q\mathcal{L}_{P}^{(0)}\tilde{\mathcal{L}}_{12}^{(0)}}, \label{seebeck}
\end{equation}
is the Seebeck coefficient of the thermocouple, corresponding to the ratio of induced electric bias from reservoirs $1$ to $2$ due to a temperature difference $\Delta T=T_0-T_P$.
\begin{equation}
\Pi_{P,12}=\frac{\mathcal{L}_{P1}^{(0)}\mathcal{L}_{P2}^{(1)}-\mathcal{L}_{P1}^{(1)}\mathcal{L}_{P2}^{(0)}}{q\mathcal{L}_{P}^{(0)}\tilde{\mathcal{L}}_{12}^{(0)}}\label{pelt}
\end{equation}
represents the Peltier coefficient of the thermocouple $\Pi_{P,12}=\frac{I_P^{(1)}}{qI_1^{(0)}}=\frac{I_P^{(1)}}{-qI_2^{(0)}}$ when $T_P=T_1=T_2$.

Notably, there is a relationship between the Peltier coefficients for the system of Figs.\ \ref{fig:modelclass} and \ref{fig:model2}, given by
\begin{equation}
	\Pi_{P,12}=\left(  \Pi_{P2}-\Pi_{P1} \right)\left(1-\frac{\mathcal{L}_{12}^{(0)}}{\tilde{\mathcal{L}}_{12}^{(0)}}\right),
 \label{eq:Peltier_3vs2}
\end{equation}
where we can see that the possibility of coherent transport from reservoirs $1$ to $2$ reduces the magnitude of the Peltier effect, as not all the current has to go through $P$.

Time-reversal symmetry \cite{LarsOns} implies $\mathcal{L}_{\alpha \beta}^{(\nu)}(\Phi)=\mathcal{L}_{\beta \alpha}^{(\nu)}(-\Phi)$, leading to the 
following relation between the Seebeck \eqref{seebeck} and Peltier \eqref{pelt} coefficients of the quantum thermocouple
\begin{equation}
S_{P,12}(\Phi)=\frac{1}{T_0}\Pi_{P,12}(-\Phi), \label{equal}
\end{equation}
in manifest resemblance with the Onsager relations for two-terminal systems \cite{Callen}.
In general, the relationship between the thermoelectric coefficients for three-terminal systems is not as straightforward \cite{chiralthermsanchezjordan}.

\section{Quantum analysis of the full nonlinear currents in the system}
To understand the full extent at which the Aharonov-Bohm effect plays a role on the thermoelectric properties of the system, the currents beyond linear response are found and are used to verify if a violation of thermodynamics such as \eqref{paradox} is present.

\subsection{Dissipative currents}
The full currents into reservoirs are studied through the formalism of non-equilibrium Green's functions (NEGF) \cite{stefanucci,jauhowingreen}. Considering a non-interacting multiterminal system, the total currents into a reservoir $\alpha$ are given by the multiterminal Büttiker-Sivan-Imry formula \cite{buttiker1,Imry1,bergnano2009},
\begin{equation}
I^{(\nu)}_\alpha=\frac{1}{h}\int d\omega \left(\omega-\mu_\alpha\right)^\nu\sum_\beta T_{\alpha\beta}\left[f_\beta-f_\alpha\right], \label{landauer}
\end{equation}
where $h$ is Planck's constant, $\nu=0$ corresponds to particle current and $\nu=1$ to heat current. $T_{\alpha\beta}(\omega)$ is the transmission function from reservoir $\beta$ to $\alpha$ and $f_\alpha(\omega)=\left(e^{\beta_\alpha
	(\omega-\mu_\alpha)}+1\right)^{-1}$ is the Fermi-Dirac distribution function of reservoir $\alpha$, and $\beta_\alpha=1/k_B T_\alpha$.
 
In the linear regime, Eq.\ \eqref{landauer} becomes Eq.\ \eqref{line1}. In this quantum mechanical description, however, the linear response coefficients may be expressed in terms of the transmission function
\begin{equation}
\mathcal{L}_{\alpha\beta}^{(\nu)}=\frac{1}{h}\int d\omega\left(\omega-\mu_{0}\right)^\nu T_{\alpha\beta}(\omega)\left(-\frac{\partial f_0}{\partial \omega}\right). \label{lincoef}
\end{equation}

We utilize Green's functions to evaluate Eq.\ \eqref{landauer} \cite{datta}
\begin{equation}
T_{\alpha\beta}(\omega)= \text{Tr}\left[ \Gamma^{\alpha}(\omega)G^R(\omega)\Gamma^{\beta}(\omega)G^A(\omega) \right],
\end{equation}
where the retarded Green's function of the system is defined as $G^R_{n\sigma,n'\sigma'}(t)=-i\theta(t)\left\langle \{d_{n\sigma}(t),d^\dagger_{n'\sigma'}(0) \}\right\rangle$, where the average is statistical and quantum. This object represents the probability amplitude of a particle in the quantum system originally in state $n'$ and with spin $\sigma'$ at time $t=0$ to be in state $n$ with a spin $\sigma$ at $t$.
$G^A(\omega)=\left[G^R(\omega)\right]^\dagger$ where $G^R(\omega)$ is the Fourier transform of $G^R(t)$. Finally, $\Gamma^\alpha(\omega)$ is the tunneling width matrix for lead $\alpha$ and describes its coupling to the system.

\subsection{Persistent currents}\label{3b}
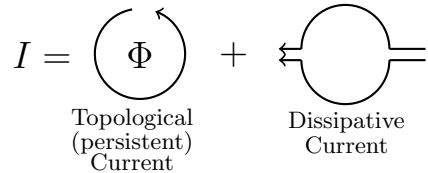
\begin{figure}[h]
	\begin{center}
		\begin{tikzpicture}[scale=1.5]
		\draw[thick,->] (0,0) arc (100:430:0.4 and 0.4);
		\node[below] at (0,-0.8) {Topological};
		\node[below] at (0,-1) {(persistent)};
        \node[below] at (0,-1.2) {Current};
		\node[] at (0.9,-0.4) {\Large{$+$}};
		\draw[thick,<-] (1.3,-0.35) -- (1.5,-0.35);
		\draw[thick] (1.5,-0.35) arc(180:0:0.39);
		\draw[thick] (2.28,-0.35) -- (2.6,-0.35);
		\draw[thick,<-] (1.3,-0.45) -- (1.5,-0.45);
		\draw[thick] (1.5,-0.45) arc(180:360:0.39);
		\draw[thick] (2.28,-0.45) -- (2.6,-0.45);
		\node at (1.9,-1) {Dissipative};
		\node at (1.9,-1.2) {Current};
		\node[] at (-0.8,-0.4) {\Large{$I=$}}; \label{currents}
		\node[] at (0.07,-0.4) {\Large{$\Phi$}}; \label{currents}
		\end{tikzpicture}
		\caption{\label{fig:currs}  %
  Representation of the different currents present in the system}
	\end{center}
\end{figure}

In addition to the dissipative currents, there is a persistent
topological current induced by the Aharonov-Bohm effect \cite{buttikerpers,AltshulerImryGeffen91} even in the absence of electric or thermal bias among the reservoirs, if magnetic flux $\Phi$ is enclosed by a path traversed by carriers as depicted in Fig.\ \ref{fig:currs}. 

For the system in equilibrium, the topological current can be expressed \cite{AltshulerImryGeffen91} in terms of the system's grand canonical potential $\Omega$ and its dependence on the magnetic flux $\Phi$ threading the system as  $I_{pers}^{(0)}=-\frac{c}{q} \frac{\partial \Omega}{\partial \Phi}$. 

An expression for the total topological electric current in the non-equilibrium steady-state system can be deduced by extending the interpretation for the equilibrium case.  For noninteracting quantum particles, the scattering states emanating from distinct reservoirs coexist in real space but are 
orthogonal in Hilbert space.  The free energy function for the nonequilibrium steady state can thus be considered as the sum of contributions that are separately in equilibrium with their reservoirs of origin
\begin{equation}
\Omega=\sum_{\alpha}\Omega_\alpha,
\end{equation} 
where
\begin{equation}
	\Omega_\alpha=-k_BT_\alpha\int d\omega \  g_\alpha\left(\omega;\Phi\right)\log\left(1+e^{-\beta_\alpha (\omega-\mu_\alpha)}\right)
 \label{eq:Omega_alpha}
\end{equation}
is the grand canonical potential of the subsystem that is in equilibrium with reservoir $\alpha$, and
\begin{equation}
    g_\alpha(\omega;\Phi)=\frac{1}{2\pi}\text{Tr}\left\lbrace G^R\Gamma^\alpha G^A \right\rbrace
\end{equation}
is 
the partial density of states of the system due to scattering states incident on it from reservoir $\alpha$ (termed injectivity in Ref.\ \cite{PDOS}). 
Note that the flux dependence in Eq.\ \eqref{eq:Omega_alpha} is carried only by the partial density of states.

Introducing the circulation frequency
\begin{equation}
\nu_\alpha(\omega;\Phi)=\frac{c\frac{\partial \chi_\alpha}{\partial \Phi}}{q g_\alpha}
\end{equation} 
for particles incident from reservoir $\alpha$,
where 
\begin{equation}
    \chi_\alpha(\omega)=\int_{-\infty}^{\omega}d\omega' g_\alpha\left(\omega';\Phi\right)
\end{equation} 
is introduced through an integration by parts (see Appendix \ref{appendix:circfreq}), the persistent particle current in this steady-state
non-equilibrium system can be expressed as
\begin{equation}
I_{pers}^{(0)}=-\frac{c}{q} \frac{\partial \Omega}{\partial \Phi}=\sum_\alpha \int d\omega \nu_\alpha g_\alpha  f_\alpha. \label{perscurr0}
\end{equation}
This equation describes the total current as being the sum of the contributions, at every energy, of the occupation of carriers incident from reservoir $\alpha$, $f_\alpha g_\alpha$, times the frequency $\nu_\alpha$ at which they encircle the AB flux, summed over all the reservoirs. 

The persistent particle current implies the existence of a persistent heat current $I_{pers}^{(1)}$ for which we propose a formula through the same reasoning as above,
\begin{equation}
I_{pers}^{(1)}=\sum_\alpha T_\alpha \int d\omega \nu_\alpha g_\alpha  s(f_\alpha), \label{perscurr1}
\end{equation}
where $s(f)=-k_B\left[f\log(f)+(1-f)\log(1-f) \right]$ is entropy of a fermionic orbital with mean occupancy $f$. Here $s(f)$ takes the role of $f$ in Eq.\ \eqref{perscurr0}, shifting the interpretation of the integrand to the entropy per state per unit energy $g_\alpha s(f_\alpha)$ times the circulation frequency of those states. The integral represents the persistent entropy current per reservoir which is then multiplied by $T_\alpha$ to yield the heat current contribution of the corresponding reservoir.

From here a quantity $\Pi_{pers}$, ``persistent Peltier coefficient'', can be defined in analogy to the Peltier coefficients for dissipative currents--
\begin{equation}
\Pi_{pers}=\frac{I^{(1)}_{pers}}{qI^{(0)}_{pers}}. \label{perspelt}
\end{equation}
The persistent Peltier coefficient $\Pi_{pers}$ is, as discussed below, a separate entity independent of the Peltier coefficients describing dissipative currents flowing between the external reservoirs $\Pi_{\alpha \beta}$.

\section{Model Quantum Thermocouple}
\label{section:Model}
A model is built to test our interpretation of the currents while also exploring to what extent the breaking of time-reversal symmetry induced by the
AB %
effect modifies thermoelectric effects in quantum systems.
We found that in this model, the AB effect can enhance and even induce thermoelectric properties when the system itself has none.

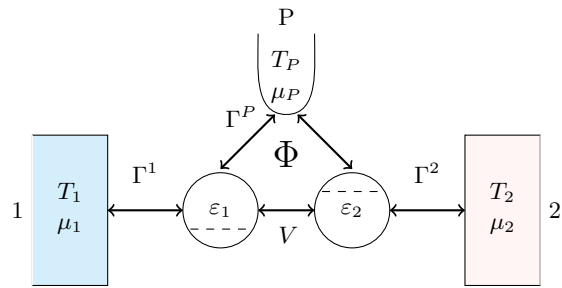
\begin{figure}[h]
\tikzset{->-/.style={decoration={
			markings,
			mark=at position .5 with {\arrow{>}}},postaction={decorate}}}
\begin{tikzpicture}[scale=0.5]
\draw[fill=cyan!15] (0,0) rectangle (2,4);
\draw[white] (0,0) -- (0,4);
\node[left] at (0,2) {$1$};
\node[above] at (1,2) {$T_{1}$};
\node[below] at (1,2) {$\mu_{1}$};
\draw[<->,thick] (2,2) -- (4,2);
\node[above] at (3,2.5) {$\Gamma^1$};
\draw[] (5,2) circle (1cm);
\draw[dashed] (4.2,1.5) -- (5.7,1.5);
\node[] at (5,2) {$\varepsilon_{1}$};
\draw[<->,thick] (6,2) -- (7.5,2);
\node[below] at (6.8,1.8) {$V$};
\draw[] (8.5,2) circle (1cm);
\draw[dashed] (7.7,2.5) -- (9.3,2.5);
\node[] at (8.5,2) {$\varepsilon_{2}$};
\draw[<->,thick] (9.5,2) -- (11.5,2);
\node[above] at (10.5,2.5) {$\Gamma^2$};
\draw[fill=pink!15] (11.5,0) rectangle (13.5,4);
\draw[white] (13.5,0) -- (13.5,4);
\node[right] at (13.5,2) {$2$};
\node[above] at (12.5,2) {$T_{2}$};
\node[below] at (12.5,2) {$\mu_{2}$};
\draw[] (6.75,4.55) to [out=180,in=270] (6.0,6.7);
\draw[] (6.75,4.55) to [out=0,in=270] (7.5,6.7);
\node[above] at (6.75,6.7) {P};
\node[above] at (6.75,5.5) {$T_{P}$};
\node[below] at (6.75,5.5) {$\mu_{P}$};
\draw[<->, thick] (5,3.1) -- (6.45,4.55);
\draw[<->, thick] (8.5,3.1) -- (7.05,4.55);
\node[] at (6.75,3.5) {\Large{$\Phi$}};
\node[] at (5.55,4.5) {$\Gamma^P$};
\end{tikzpicture}
\vspace{0.3cm}
	\caption{\label{fig:model1}  %
 Schematic representation of the model quantum thermocouple.}
\end{figure}

The system is comprised of two quantum dots as shown in Fig.\ \ref{fig:model1}, with resonances $\varepsilon_{1,2}$. The quantum system is coupled to three terminals and it is mainly studied for the conditions in which it functions as a quantum thermocouple ($\mu_2>\mu_1$, $T_1=T_2=T_P$.)

For the system to be able to work as a thermocouple, one of the terminals, labeled as $P$, must have a chemical potential $\mu_{P}$ tuned such that it is electrically floating, i.e., $I^{(0)}_P=0 $.

\subsection{Hamiltonian}
The model Hamiltonian may be written as
\begin{equation}
\hat{H}=	\hat{H}_{\rm sys}+	\hat{H}_{\rm res}+	\hat{H}_{\rm sr},
\end{equation}
where %
\begin{equation}
 \hat{H}_{\rm sys}=\sum_{i=1,2} \varepsilon_{i}d_i^\dagger d_i+Vd_2^\dagger d_1+V^* d_1^\dagger d_2\label{hsys}
\end{equation}
describes the isolated double-dot system,
\begin{equation}
\hat{H}_{\rm res}=\sum_{\alpha=1,2,P}\,\sum_{k\in \alpha} \varepsilon_{k} c_{k}^\dagger c_{k}
\end{equation}
describes the three macroscopic terminals (reservoirs), and
\begin{equation}
\hat{H}_{\rm sr}=\sum_{n=1}^2\sum_{\alpha=1,2,P}\sum_{k\in \alpha} V_{kn}(\Phi)c_{k}^\dagger d_n+\mbox{h.c.}
\end{equation}
describes the coupling of the double-dot to the three external terminals.
Here $d_n^\dagger$ is the fermionic operator that creates a particle in the $\varepsilon_n$ level of the quantum system and $c_{k}^\dagger|_{k\in\alpha}$ creates a fermionic particle in state $k$ of reservoir $\alpha$. The spin index of the creation and annihilation operators has been suppressed \cite{footnote_spin}.
The levels of the dots are arranged such that one of them behaves as a donor of electrons (dot 2) and the other as an acceptor (dot 1).

A magnetic flux $\Phi$ is threaded through the system, in order to produce a persistent current through the Aharonov-Bohm effect. The flux $\Phi$ is set and contained in a closed
loop of transit of electrons from the dots to the
probe $P$ and in between the dots (see Figs.\ \ref{fig:model2}, \ref{fig:model1}), while the magnetic field within the double-dot itself and the terminals is zero. The influence of $\Phi$ on the system is thus a pure quantum effect; there is no classical Lorentz force.

The tunneling matrix elements $V_{kn}(\Phi)$ are related to the tunneling-width matrices $\Gamma^\alpha$ describing the junctions formed between the double-dot and the three terminals by
\begin{equation}
\left[\Gamma^\alpha(\Phi;\omega)\right]_{nn'}=2\pi\sum_{k\in \alpha}V_{nk}(\Phi)V_{n'k}^*(\Phi)\delta(\omega-\varepsilon_{k}).
\end{equation}
In the following, we will assume the coupling to the terminals can be described in the broad-band limit \cite{Shastry2016}, where $\Gamma^\alpha(\Phi)$ is independent
of the energy $\omega$.
A choice of gauge \cite{byersyang} can be made such that the effect of the magnetic flux $\Phi$ threading the system is encoded only in $\Gamma^P$ through
\begin{equation}
\Gamma^P(\Phi)=\gamma_p 
\begin{pmatrix}
1 & e^{i2\pi  \frac{\Phi}{\phi_0}} \\
e^{-i2\pi  \frac{\Phi}{\phi_0}} & 1
\end{pmatrix},
\end{equation}
where $\phi_0=\frac{hc}{q}$ is the magnetic flux quantum.

The dissipative currents into the external terminals were calculated numerically by evaluation of Eq.\ \eqref{landauer}, while the persistent currents flowing within the system were calculated using Eqs.\ \eqref{perscurr0}, \eqref{perscurr1}, and \eqref{perspelt}. A resolution of the paradox introduced in section \ref{section:paradox} is attained by a physical analysis of the nature of the persistent current as a separate entity from the dissipative currents.

Due to the number of parameters in the model, the presentation of the results is not exhaustive, but the conclusions drawn take into account an evaluation over a broad set of configurations in parameter space.

\subsection{Results for the external and persistent currents}
\begin{figure}[h]
	\includegraphics[width=1\linewidth]{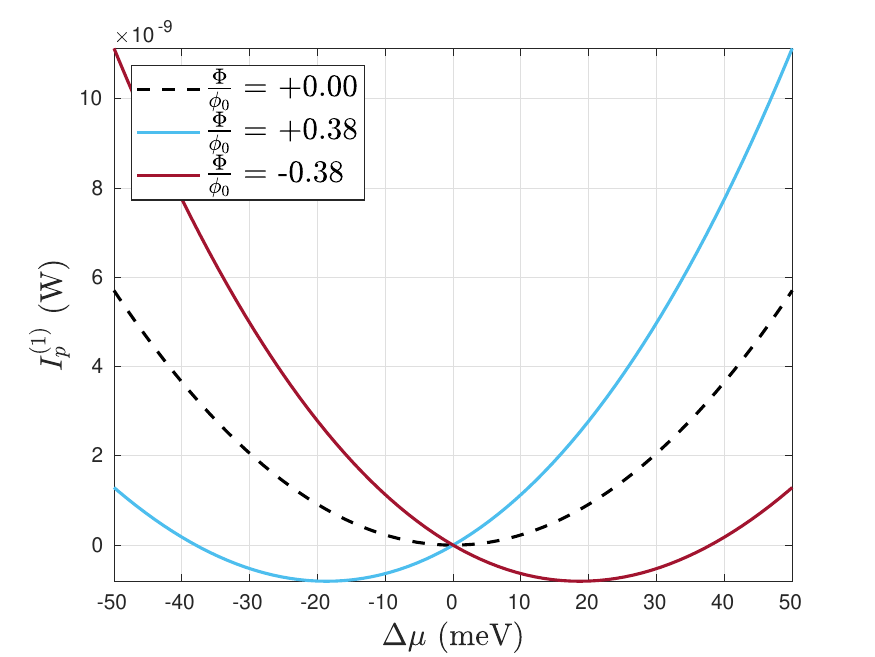}
	\caption{Joule heating and Peltier cooling/heating in a model quantum thermocouple. Here $\varepsilon_1=\varepsilon_2=0$, $\gamma_\alpha=0.1$ eV$\ \forall \alpha$, $V=0.1$ eV, $T_\alpha=300K$ $\forall$ $\alpha$, and $\Delta \mu=\mu_{2}-\mu_{1} $, with the Fermi level $\mu_0=0$. %
 Maximum probe cooling on the order of 1nW is obtained in optimal conditions in the model quantum thermocouple. At $\Phi=0$, the system  exhibits no thermoelectric response due to particle-hole symmetry ($\varepsilon_1=\varepsilon_2$). }\label{fig:heat}
\end{figure}

Direct evaluation of Eqs.\ \eqref{landauer}, \eqref{lincoef}, \eqref{perscurr0}, and \eqref{perscurr1}
showed that the thermoelectric effects induced by the AB effect are significant even when there are no n-type or p-type elements in the junction, as shown in Fig.\ \ref{fig:heat}. With the parameters specified in  Fig.\ \ref{fig:heat}, cooling of the order of $\sim$1 nW was achieved, in a system that originally lacked thermoelectric properties.

\begin{figure}[h]
	\includegraphics[width=1\linewidth]{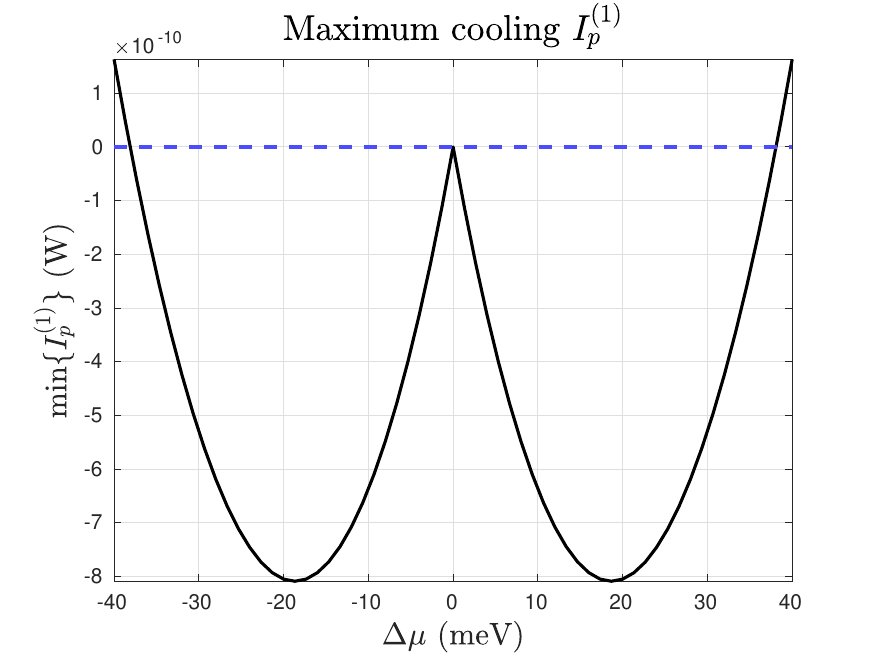}
	\caption{Maximum AB cooling of the probe in the particle-hole symmetric system ($\varepsilon_1=\varepsilon_2$) as a function of the electric bias.  The parameters of the quantum thermocouple are given in Fig.\ \ref{fig:heat}.}\label{fig:maxcool} %
\end{figure}

Fig.\ \ref{fig:maxcool} shows the maximum achievable probe cooling from the AB effect in the model quantum thermocouple as a function of the source-drain bias $\Delta\mu$.
As in a classical thermocouple (cf.\ Fig.\ \ref{fig:modelclass}), there is an upper limit to the Peltier cooling since the Peltier effect is linear and Joule heating is quadratic in bias.

\begin{figure}%
	\includegraphics[width=1\linewidth]{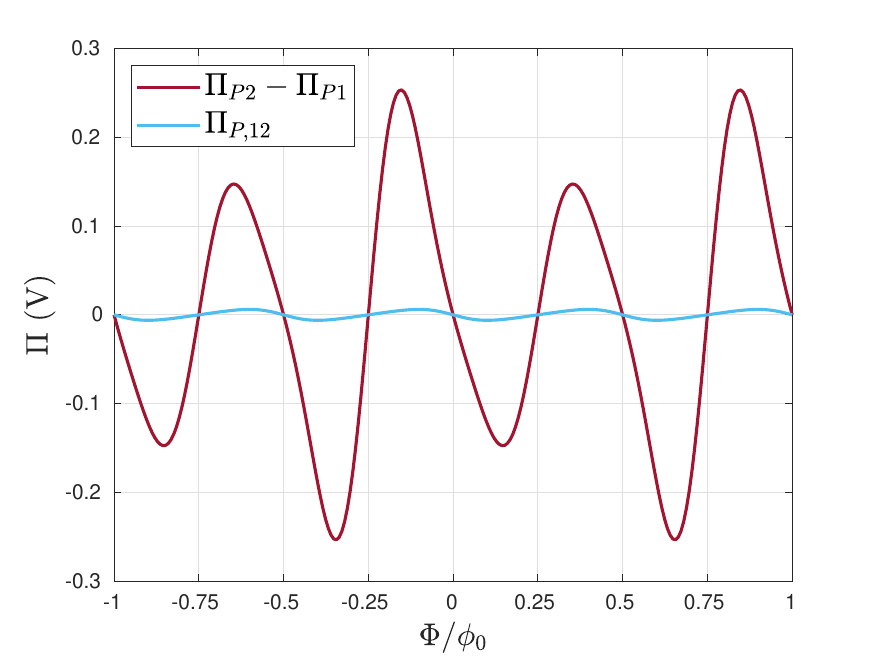} %
	\caption{Comparison of the effective Peltier coefficient of the three-terminal quantum system $\Pi_{P,12}$ and difference in 2-terminal Peltier coefficients $\Pi_{P2}-\Pi_{P1}$. The model parameters are given in Fig.\ \ref{fig:heat}.
 }\label{fig:comppis} %
\end{figure}

The Peltier coefficients for the model quantum thermocouple are shown as functions of flux in Fig.\ \ref{fig:comppis}. The system acquires thermoelectric properties by turning on the AB flux $\Phi$; it can be set to work as a thermocouple solely by induction of the Aharonov-Bohm effect, a fully quantum effect. It is also shown that the effective 3-terminal Peltier coefficent $\Pi_{P,12}$ %
is in general smaller in magnitude than the corresponding Peltier coefficient for a classical thermocouple $\Pi_{P2}-\Pi_{P1}$, consistent with Eq.\ \eqref{eq:Peltier_3vs2}. This is expected, as the introduction of a direct transport channel from 1 to 2 reduces the overall share of particles traversing reservoir $P$, and consequently reduces the ratio of heat into $P$ to the source-drain current.  Nonetheless, the direct path is needed in the quantum thermocouple to provide a circuit encircling the AB flux.

\begin{figure}%
    \includegraphics[width=1\linewidth]{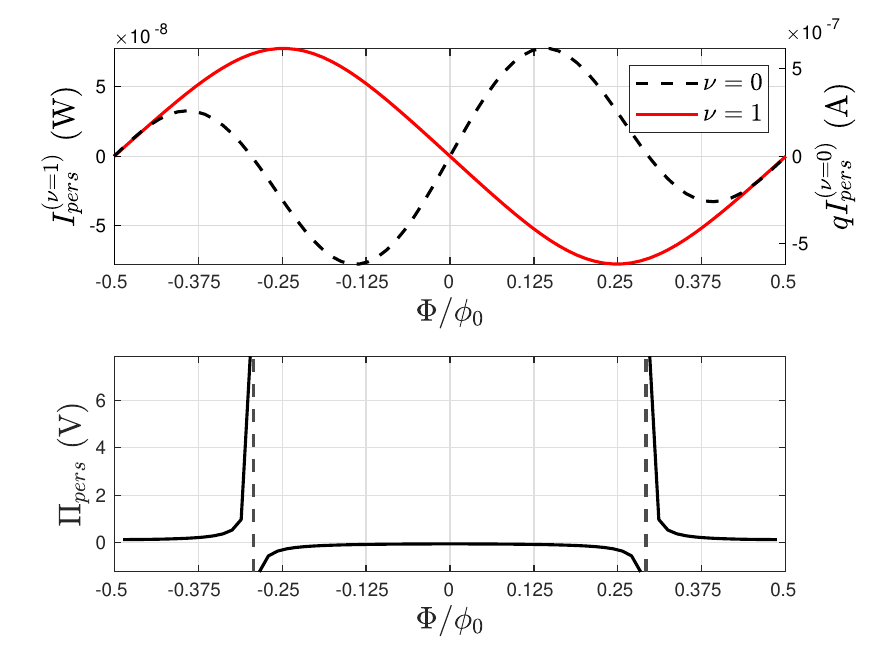}
	\caption{ Top: Persistent currents defined by Eqs.\ \eqref{perscurr0} and \eqref{perscurr1} as functions of the AB flux $\Phi$ for a quantum thermocouple with particle-hole symmetry ($\varepsilon_1=\varepsilon_2$) at $\Delta \mu = 0$. 
 The model parameters are given in Fig.\ \ref{fig:heat}.
 Persistent currents of heat are present even in the absence of electric or thermal bias. Bottom: Persistent Peltier coefficient defined in Eq.\ \eqref{perspelt}}\label{fig:pers} %
\end{figure}

It was observed that the magnitude of the persistent heat current can be comparable to or greater than the total (dissipative) heat current into $P$ (see Fig.\ \ref{fig:pers}). This highlights the importance of the persistent currents, as they play a significant role in the overall %
transport of particles and entropy within the system.

A comparison of Figs.\ \ref{fig:comppis} and \ref{fig:pers} shows that
the persistent Peltier coefficient defined by Eq.\ \eqref{perspelt} can be significantly larger than the classical Peltier coefficient $\Pi_{P2}-\Pi_{P1}$ of the thermocouple, particularly in the vicinity of the resonances evident in Fig.\ \ref{fig:pers}, where $\Pi_{\rm pers}\rightarrow \pm \infty$ (not shown). It is not known whether the divergence of the persistent Peltier coefficient is physically significant, but it is conceivable that such an effect could be used to control the flow of heat in a quantum circuit.  This enhancement of the Peltier effect on the persistent current is a quantum interference effect somewhat analogous to the enhancement of the Seebeck effect in the vicinity of a transmission node \cite{bergnano2009}.

\subsection{Resolution of the paradox}

\begin{figure}%
\includegraphics[width=1\linewidth]{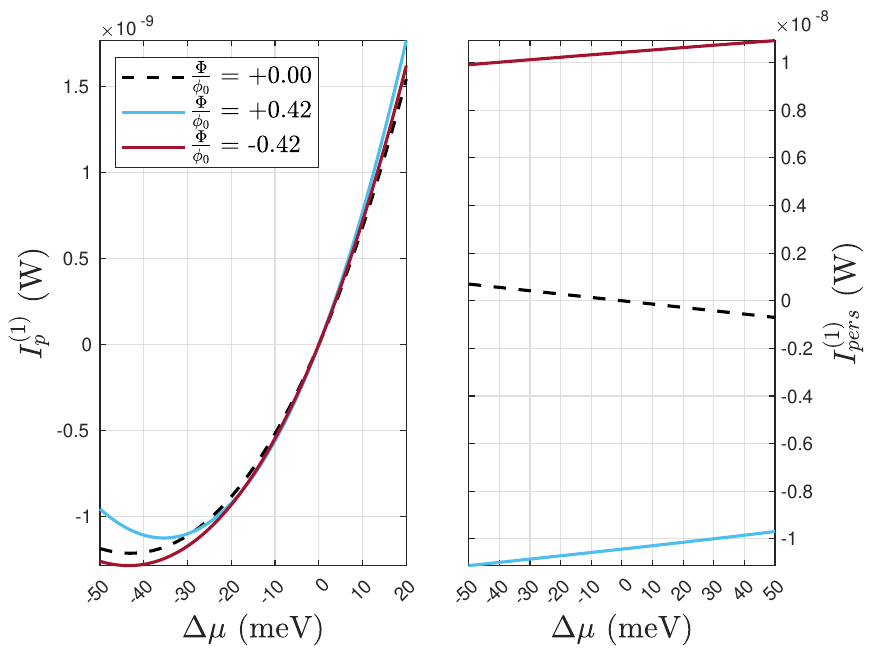}
	\caption{Left: total current into P. Right: persistent current. A system with built-in thermoelectric properties does not exhibit cooling when there is no electric bias, even when a persistent heat current (of comparable, and in this case, higher magnitude than the total current into P) is present due to the AB effect. The model parameters are the same as the ones given in Fig.\ \ref{fig:heat}, except $\varepsilon_2-\varepsilon_1=0.5\mbox{eV}$.
 }\label{fig:heatbias} %
\end{figure}

\begin{figure}%
\includegraphics[width=1\linewidth]{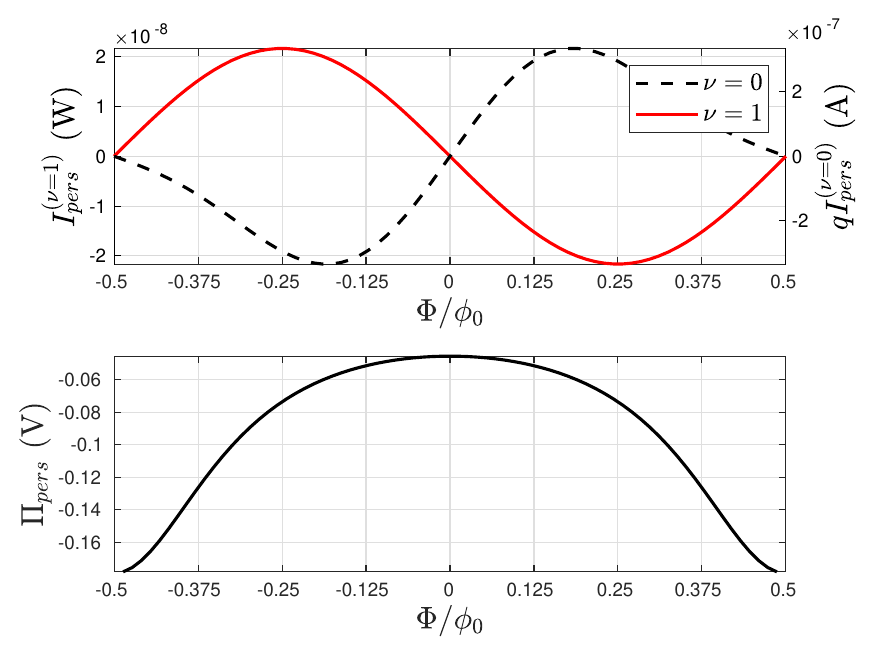}
	\caption{Top: Persistent currents are present within the quantum thermocouple of Fig.\ \ref{fig:heatbias} ($\varepsilon_1-\varepsilon_2=0.5$ eV) even in the absence of electric or thermal bias ($\Delta \mu=0$, $\Delta T=0$), but they represent circulating currents of entropy and particles that are non-dissipative, and does not lead to cooling or heating of the probe. Bottom: Persistent peltier coefficient with these parameters.}\label{fig:persbias} 
\end{figure}

Consider now a quantum system configured with $\varepsilon_1\neq \varepsilon_2$ 
so that it functions as a %
thermocouple with $\Pi_{P,12}\neq 0$ even at $\Phi=0$.  Based on the arguments of Sec.\ \ref{section:paradox},
one might expect a persistent electric current induced in the system by an AB flux $\Phi$ to lead to persistent cooling or heating of the probe according to Eq.\ \eqref{paradox}.
However, a direct calculation using Eq.\ \eqref{landauer} reveals that there is no net cooling of the probe at $\Delta \mu=0$ regardless of the value of the AB flux $\Phi$, as shown in Fig.\ \ref{fig:heatbias}.  It is worth noticing that in Fig.\ \ref{fig:heatbias}, the persistent current may be of comparable magnitude to the dissipative current, and that they strongly depend on the flux and only weakly on the bias.  Even though there are generically persistent currents present in the system when $\Phi\neq 0$, as shown in Fig.\ \ref{fig:persbias}, which depict a net flow of charge carriers through reservoir $P$, no net cooling takes place when $\Delta\mu=0$. 

The thermoelectric properties of the system are still modified by the AB effect, as in the particle-hole symmetric system: the intrinsic
Peltier effect at $\Phi=0$ can be either enhanced or suppressed, depending on the value of $\Phi$ (see Fig.\ \ref{fig:heatbias}).
 This is because of the shift in the energy levels in the quantum system due to the magnetic flux. Effectively, $\Phi$ makes the ``p-type" quantum dot more ``p-type" by further reducing $\varepsilon_1$ with respect to the other level and the ``n-type" dot more ``n-type" by raising $\varepsilon_2$, increasing the magnitude of the effective Peltier coefficient in one direction of the flux (in this case when $\Phi<0$). If the flux is reversed, the opposite effect can also be observed.

\begin{figure}%
    \resizebox{1\linewidth}{!}{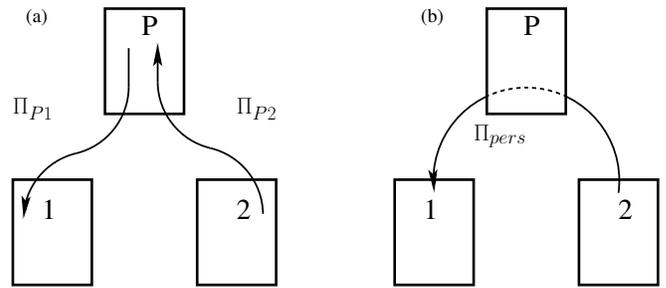}
	\caption{(a) Schematic diagram of currents in a floating probe. The flow is incoherent and sequential, since it involves two individual quantum processes, from 2 \textit{into} $P$ and from $P$ into 1. (b) Persistent current involves waves that virtually enter $P$, this is a one-step quantum process and is coherent. %
 }\label{fig:crrc} %
\end{figure}

 The persistent cooling paradox arises because of a superficial similarity of the persistent electric current that traverses the probe in the AB configuration of Figs.\ \ref{fig:model2} and \ref{fig:model1} and the dissipative electric current that traverses the floating electric probe under finite source-drain bias, as shown, for example, in Fig.\ \ref{fig:modelclass}. The electric current traversing the floating probe at finite source-drain bias is the incoherent sum of currents from the source and drain, which must add to zero if the probe is an open electric terminal \cite{PhysRevB.40.3409,5390027} [see Fig.\ \ref{fig:crrc}(a)].  However, the equilibrium electric current traversing the probe induced by the AB effect represents a single conserved quantum coherent flux; the currents in and out are equal by necessity, not due to a floating condition [see Fig.\ \ref{fig:crrc}(b)].  Corresponding to the different character of the electric current through the probe in these two scenarios, the Peltier effect is also different in character.  In the first case, the incoherent electric currents from the source and drain entrain separate heat currents according to Eq.\ \eqref{piclas} that by design do not cancel, while in the second case, the persistent electric current entrains a persistent heat current through the probe according to Eq.\ \eqref{perspelt}, a conserved quantum coherent flux such that the heat currents into and out of the probe are equal by necessity.

 The paradox is resolved by taking into account the different character of dissipative versus equilibrium currents, and the corresponding difference between the dissipative and equilibrium Peltier effects.  The equilibrium, or persistent, Peltier effect gives rise to a conserved heat current flowing within the system, and causes no cooling or heating of any part of the circuit.  It simply corresponds to the flow of entropy entrained by the persistent flow of the charge carriers times the absolute temperature of the equilibrium system.

\section{Summary}

The basic element in thermoelectrics is the thermocouple, a three-terminal device with electric source and drain terminals and a floating electric terminal that can be either heated or cooled by the Peltier effect.  A classical thermocouple consists of two junctions in series, but a quantum thermocouple can also include a direct electrical path between source and drain, leading to a multiply connected topology which can be threaded by a magnetic flux, leading to the Aharonov-Bohm effect.

The Aharonov-Bohm (AB) effect on the thermoelectric response of a topological quantum thermocouple was investigated, and general expressions were derived using NEGF. %
Relations between the Peltier and Seebeck coefficients in quantum thermocouples following from time-reversal symmetry and the 2nd Law of Thermodynamics were derived and interpreted.

In addition to modifying the external electric and thermal currents of the device, the AB effect also induces persistent electric and thermal currents within the quantum system.
New NEGF expressions for persistent electric and thermal currents for non-interacting open quantum systems in steady state were derived, and applied to analyze a topological quantum thermocouple. %

For a model quantum thermocouple, the AB effect was shown to modify its thermoelectric properties, endowing the system with large thermoelectric effects where nominally none should occur (electron-hole symmetric case). The Peltier effect in a system of such nature can only be understood as a purely quantum mechanical effect.
The topological thermocouple also exhibits a novel type of Peltier effect linking its persistent thermal and electric currents.

In a classical thermocouple, an electric current flowing sequentially from the source to the drain through the third terminal leads to Peltier cooling or heating of that terminal, a solid-state realization of a heat pump.  
One might expect that a persistent electric current flowing through the third terminal of a topological quantum thermocouple
could lead to persistent Peltier cooling of that terminal.  This would, however, violate both the 1st and 2nd Laws of Thermodynamics, and is shown not to occur.   This apparent paradox is resolved by elucidating the distinction between persistent and dissipative currents in quantum thermoelectrics.  

A re-interpretation of the effect of persistent currents (that are observed to be of comparable magnitude to dissipative currents) over the reservoirs has to be introduced in order to avoid the thermodynamic paradox. Dissipative and persistent currents present must be acknowledged as phenomena of a different nature. In a topological quantum thermocouple, persistent currents do not contribute to total heat or particle currents into any reservoir, but traverse them without losing phase coherence, and in general they represent only conserved fluxes of energy and particles in the system, which arise from the boundary conditions (in this case the electrical work needed to set the persistent current).

\section{Author contributions}
Y.X.\ exclusively worked on the mathematical framework for persistent particle and entropy currents in equilibrium systems discussed in Appendix \ref{appendix:circfreq}, that was extended to the nonequilibrium steady state in Sec.\ \ref{3b}. M.J-V.\ and C.A.S.\ are responsible for the rest of the paper. 

\begin{acknowledgments}
We are grateful for the insightful discussions with Ferdinand Evers, Parth Kumar, and Caleb M. Webb. This work was partially supported by the U.S. Department of Energy (DOE), Office of Science, under Award No. DE-SC0006699.
\end{acknowledgments}

\appendix

\section{Circulation frequency in the open system in equilibrium}
\label{appendix:circfreq}

In the equilibrium case, an open system coupled to a single reservoir is described by a grand canonical potential
\begin{equation}
    \Omega=-k_BT\int d\omega \ g\left(\omega;\Phi\right)\log\left( 1+e^{-\beta\left(\omega-\mu\right)} \right),
\end{equation}
where $g\left(\omega;\Phi\right)=\frac{1}{2\pi}\text{Tr}\left\lbrace G^R\Gamma G^A \right\rbrace$, which in this case with a single reservoir corresponds to the density of states of the system, considering the broadening \cite{datta} of the levels brought about by the connection of the system to the reservoir described by $\Gamma$.

Using the known relation described in Sec.\ \ref{3b}
\begin{equation}
    I_{pers}^{(0)}=-\frac{c}{q}\frac{\partial \Omega}{\partial \Phi},
\end{equation}
and performing the partial derivative
\begin{equation}
    \frac{\partial \Omega}{\partial \Phi}=-k_BT\int d\omega \frac{\partial g\left(\omega;\Phi\right)}{\partial \Phi}\log\left( 1+e^{-\beta\left(\omega-\mu\right)} \right),
\end{equation}
where it can be explicitly seen that all the $\Phi$ dependence is carried by $g$. Performing an integration by parts, and noting that the boundary terms vanish since the spectrum is bounded below and the statistical factor vanishes as $\omega\rightarrow +\infty$, one obtains
\begin{equation}
    \frac{\partial \Omega}{\partial \Phi} =-\int d\omega\  \frac{\partial \chi\left(\omega;\Phi\right)}{\partial \Phi} f(\omega),
\end{equation}
where $f$ is the usual Fermi distribution function and
\begin{equation}
\chi\left(\omega;\Phi\right)=\int_{-\infty}^{\omega} d\omega'\ g\left(\omega';\Phi\right).
\end{equation}

The persistent particle current is then
\begin{equation}
      I_{pers}^{(0)}=\int d\omega\ \nu\left(\omega;\Phi\right) g\left(\omega;\Phi\right)f\left(\omega\right),   
\end{equation}
where
\begin{equation}
 \nu\left(\omega;\Phi\right) \equiv \frac{c}{q}\frac{1}{g\left(\omega;\Phi\right)}\frac{\partial \chi}{\partial \Phi}
\end{equation}
is the circulation frequency associated with states of energy $\omega$, that is, the mean rate to encircle the AB flux as a function of energy, as dimensional analysis will also confirm.

The extension to the non-equilibrium steady state for a system of fermions without interparticle interactions leads directly to 
Eq.\ \eqref{perscurr0} 
since the contributions of the distinct subsystems separately in equilibrium with each reservoir are simply additive in the absence of interparticle interactions.

\bibliographystyle{plainnat}
\bibliography{persvdissiparxiv.bib}

\providecommand{\noopsort}[1]{}\providecommand{\singleletter}[1]{#1}%
\begin{thebibliography}{35}
\providecommand{\natexlab}[1]{#1}
\providecommand{\url}[1]{\texttt{#1}}
\expandafter\ifx\csname urlstyle\endcsname\relax
  \providecommand{\doi}[1]{doi: #1}\else
  \providecommand{\doi}{doi: \begingroup \urlstyle{rm}\Url}\fi

\bibitem[foo()]{footnote_spin}
The system is assumed to possess spin-rotation symmetry and spin-orbit coupling has been neglected.

\bibitem[Altshuler et~al.(1991)Altshuler, Gefen, and Imry]{AltshulerImryGeffen91}
B.~L. Altshuler, Y.~Gefen, and Y.~Imry.
\newblock Persistent differences between canonical and grand canonical averages in mesoscopic ensembles: Large paramagnetic orbital susceptibilities.
\newblock \emph{Phys. Rev. Lett.}, 66:\penalty0 88--91, Jan 1991.
\newblock \doi{10.1103/PhysRevLett.66.88}.
\newblock URL \url{https://link.aps.org/doi/10.1103/PhysRevLett.66.88}.

\bibitem[Antti-Pekka~Jauho(1994)]{jauhowingreen}
Yigal~Meir Antti-Pekka~Jauho, Ned S.~Wingreen.
\newblock \emph{Phys. Rev. B}, 50:\penalty0 5528, 1994.

\bibitem[Bergfield and Stafford(2009)]{bergnano2009}
J.~P. Bergfield and C.~A. Stafford.
\newblock Thermoelectric signatures of coherent transport in single-molecule heterojunctions.
\newblock \emph{Nano Letters}, 9\penalty0 (8):\penalty0 3072--3076, 2009.
\newblock \doi{10.1021/nl901554s}.
\newblock URL \url{https://doi.org/10.1021/nl901554s}.
\newblock PMID: 19610653.

\bibitem[Bergfield et~al.(2010)Bergfield, Solis, and Stafford]{Bergfield2010}
Justin~P. Bergfield, Michelle~A. Solis, and Charles~A. Stafford.
\newblock Giant thermoelectric effect from transmission supernodes.
\newblock \emph{ACS Nano}, 4\penalty0 (9):\penalty0 5314--5320, 2010.
\newblock \doi{10.1021/nn100490g}.
\newblock URL \url{https://doi.org/10.1021/nn100490g}.
\newblock PMID: 20735063.

\bibitem[B{\'e}rut et~al.(2012)B{\'e}rut, Arakelyan, Petrosyan, Ciliberto, Dillenschneider, and Lutz]{berutExperimentalVerificationLandauers2012}
Antoine B{\'e}rut, Artak Arakelyan, Artyom Petrosyan, Sergio Ciliberto, Raoul Dillenschneider, and Eric Lutz.
\newblock Experimental verification of {{Landauer}}'s principle linking information and thermodynamics.
\newblock \emph{Nature}, 483\penalty0 (7388):\penalty0 187---189, March 2012.
\newblock ISSN 0028-0836.
\newblock \doi{10.1038/nature10872}.
\newblock URL \url{https://doi.org/10.1038/nature10872}.

\bibitem[B\"uttiker(1986)]{buttiker1}
M.~B\"uttiker.
\newblock Four-terminal phase-coherent conductance.
\newblock \emph{Phys. Rev. Lett.}, 57:\penalty0 1761--1764, Oct 1986.
\newblock \doi{10.1103/PhysRevLett.57.1761}.
\newblock URL \url{https://link.aps.org/doi/10.1103/PhysRevLett.57.1761}.

\bibitem[Buttiker(1988)]{5390027}
M.~Buttiker.
\newblock Coherent and sequential tunneling in series barriers.
\newblock \emph{IBM Journal of Research and Development}, 32\penalty0 (1):\penalty0 63--75, 1988.
\newblock \doi{10.1147/rd.321.0063}.

\bibitem[B\"uttiker(1989)]{PhysRevB.40.3409}
M.~B\"uttiker.
\newblock Chemical potential oscillations near a barrier in the presence of transport.
\newblock \emph{Phys. Rev. B}, 40:\penalty0 3409--3412, Aug 1989.
\newblock \doi{10.1103/PhysRevB.40.3409}.
\newblock URL \url{https://link.aps.org/doi/10.1103/PhysRevB.40.3409}.

\bibitem[Byers and Yang(1961)]{byersyang}
N.~Byers and C.~N. Yang.
\newblock \emph{Phys. Rev. Lett.}, 7:\penalty0 46--49, Jul 1961.
\newblock \doi{10.1103/PhysRevLett.7.46}.
\newblock URL \url{https://link.aps.org/doi/10.1103/PhysRevLett.7.46}.

\bibitem[Büttiker et~al.(1983)Büttiker, Imry, and Landauer]{buttikerpers}
M.~Büttiker, Y.~Imry, and R.~Landauer.
\newblock Josephson behavior in small normal one-dimensional rings.
\newblock \emph{Physics Letters A}, 96\penalty0 (7):\penalty0 365--367, 1983.
\newblock ISSN 0375-9601.
\newblock \doi{https://doi.org/10.1016/0375-9601(83)90011-7}.
\newblock URL \url{https://www.sciencedirect.com/science/article/pii/0375960183900117}.

\bibitem[Calder and Sondheimer(1966)]{chemists2}
I.~C. Calder and F.~Sondheimer.
\newblock \emph{Chem. Commun. (London)}, pages 904--905, 1966.
\newblock \doi{10.1039/C19660000904}.
\newblock URL \url{http://dx.doi.org/10.1039/C19660000904}.

\bibitem[Callen(1985)]{Callen}
Herbert~B Callen.
\newblock \emph{{Thermodynamics and an introduction to thermostatistics; 2nd ed.}}
\newblock Wiley, New York, NY, 1985.

\bibitem[Datta(1997)]{datta}
Supriyo Datta.
\newblock \emph{Electronic transport in mesoscopic systems}.
\newblock Cambridge university press, 1997.

\bibitem[Gasparian et~al.(1996)Gasparian, Christen, and B\"uttiker]{PDOS}
V.~Gasparian, T.~Christen, and M.~B\"uttiker.
\newblock Partial densities of states, scattering matrices, and green's functions.
\newblock \emph{Phys. Rev. A}, 54:\penalty0 4022--4031, Nov 1996.
\newblock \doi{10.1103/PhysRevA.54.4022}.
\newblock URL \url{https://link.aps.org/doi/10.1103/PhysRevA.54.4022}.

\bibitem[Goold et~al.(2016)Goold, Huber, Riera, del Rio, and Skrzypczyk]{Goold_2016}
John Goold, Marcus Huber, Arnau Riera, Lídia del Rio, and Paul Skrzypczyk.
\newblock The role of quantum information in thermodynamics—a topical review.
\newblock \emph{Journal of Physics A: Mathematical and Theoretical}, 49\penalty0 (14):\penalty0 143001, feb 2016.
\newblock \doi{10.1088/1751-8113/49/14/143001}.
\newblock URL \url{https://dx.doi.org/10.1088/1751-8113/49/14/143001}.

\bibitem[Jiang and Imry(2017)]{jiangPRApp2017}
Jian-Hua Jiang and Yoseph Imry.
\newblock Enhancing thermoelectric performance using nonlinear transport effects.
\newblock \emph{Phys. Rev. Applied}, 7:\penalty0 064001, Jun 2017.
\newblock \doi{10.1103/PhysRevApplied.7.064001}.
\newblock URL \url{https://link.aps.org/doi/10.1103/PhysRevApplied.7.064001}.

\bibitem[Koski et~al.(2014)Koski, Maisi, Pekola, and {Dmitri V. Averin}]{koskiExperimentalRealizationSzilard2014}
Jonne~V. Koski, Ville~F. Maisi, Jukka~P. Pekola, and {Dmitri V. Averin}.
\newblock Experimental realization of a {{Szilard}} engine with a single electron.
\newblock \emph{Proceedings of the National Academy of Sciences}, 111\penalty0 (38):\penalty0 13786--13789, 2014.
\newblock \doi{10.1073/pnas.1406966111}.

\bibitem[Liu et~al.(2021)Liu, Jung, and Segal]{liuPeriodicallyDrivenQuantum2021}
Junjie Liu, Kenneth~A. Jung, and Dvira Segal.
\newblock Periodically {{Driven Quantum Thermal Machines}} from {{Warming}} up to {{Limit Cycle}}.
\newblock \emph{Physical Review Letters}, 127\penalty0 (20):\penalty0 200602, November 2021.
\newblock ISSN 0031-9007, 1079-7114.
\newblock \doi{10.1103/PhysRevLett.127.200602}.
\newblock URL \url{https://link.aps.org/doi/10.1103/PhysRevLett.127.200602}.

\bibitem[Nimmagadda et~al.(2021)Nimmagadda, Mahmud, and Sinha]{microheatrev}
Lakshmi~Amulya Nimmagadda, Rifat Mahmud, and Sanjiv Sinha.
\newblock Materials and devices for on-chip and off-chip peltier cooling: A review.
\newblock \emph{IEEE Transactions on Components, Packaging and Manufacturing Technology}, 11\penalty0 (8):\penalty0 1267--1281, 2021.
\newblock \doi{10.1109/TCPMT.2021.3095048}.

\bibitem[Onsager(1931)]{LarsOns}
Lars Onsager.
\newblock Reciprocal relations in irreversible processes. i.
\newblock \emph{Phys. Rev.}, 37:\penalty0 405--426, Feb 1931.
\newblock \doi{10.1103/PhysRev.37.405}.
\newblock URL \url{https://link.aps.org/doi/10.1103/PhysRev.37.405}.

\bibitem[Pendry(1983)]{pendry1983}
J~B Pendry.
\newblock Quantum limits to the flow of information and entropy.
\newblock \emph{Journal of Physics A: Mathematical and General}, 16\penalty0 (10):\penalty0 2161--2171, jul 1983.
\newblock \doi{10.1088/0305-4470/16/10/012}.
\newblock URL \url{https://doi.org/10.1088/0305-4470/16/10/012}.

\bibitem[Riffat and Ma(2004)]{pelthist}
S.~Riffat and Xiaoli Ma.
\newblock Improving the coefficient of performance of thermoelectric cooling systems: A review.
\newblock \emph{International Journal of Energy Research}, 28:\penalty0 753 -- 768, 07 2004.
\newblock \doi{10.1002/er.991}.

\bibitem[Saito et~al.(2011)Saito, Benenti, Casati, and Prosen]{Prosen}
Keiji Saito, Giuliano Benenti, Giulio Casati, and Toma\ifmmode \check{z}\else~\v{z}\fi{} Prosen.
\newblock Thermopower with broken time-reversal symmetry.
\newblock \emph{Phys. Rev. B}, 84:\penalty0 201306, Nov 2011.
\newblock \doi{10.1103/PhysRevB.84.201306}.
\newblock URL \url{https://link.aps.org/doi/10.1103/PhysRevB.84.201306}.

\bibitem[S\'anchez et~al.(2015)S\'anchez, Sothmann, and Jordan]{chiralthermsanchezjordan}
Rafael S\'anchez, Bj\"orn Sothmann, and Andrew~N. Jordan.
\newblock Chiral thermoelectrics with quantum hall edge states.
\newblock \emph{Phys. Rev. Lett.}, 114:\penalty0 146801, Apr 2015.
\newblock \doi{10.1103/PhysRevLett.114.146801}.
\newblock URL \url{https://link.aps.org/doi/10.1103/PhysRevLett.114.146801}.

\bibitem[Seebeck(1822)]{seebeck}
T.~Seebeck.
\newblock Magnetic polarization of metals and ores by temperature differences.
\newblock \emph{Abhandlungen der Königlichen Akademie der Wissenschaften zu Berlin}, pages 265--373, 1822.

\bibitem[Seifert(2012)]{Seifert_2012}
Udo Seifert.
\newblock Stochastic thermodynamics, fluctuation theorems and molecular machines.
\newblock \emph{Reports on Progress in Physics}, 75\penalty0 (12):\penalty0 126001, nov 2012.
\newblock \doi{10.1088/0034-4885/75/12/126001}.
\newblock URL \url{https://dx.doi.org/10.1088/0034-4885/75/12/126001}.

\bibitem[Shastry and Stafford(2016)]{Shastry2016}
Abhay Shastry and Charles~A. Stafford.
\newblock Temperature and voltage measurement in quantum systems far from equilibrium.
\newblock \emph{Phys. Rev. B}, 94:\penalty0 155433, Oct 2016.
\newblock \doi{10.1103/PhysRevB.94.155433}.
\newblock URL \url{https://link.aps.org/doi/10.1103/PhysRevB.94.155433}.

\bibitem[Shroeder and Oth(1966)]{chemists1}
G~Shroeder and F.J.M. Oth.
\newblock \emph{Tetrahedron Letts.}, page 4083, 1966.

\bibitem[Sivan and Imry(1986)]{Imry1}
U.~Sivan and Y.~Imry.
\newblock Multichannel landauer formula for thermoelectric transport with application to thermopower near the mobility edge.
\newblock \emph{Phys. Rev. B}, 33:\penalty0 551--558, Jan 1986.
\newblock \doi{10.1103/PhysRevB.33.551}.
\newblock URL \url{https://link.aps.org/doi/10.1103/PhysRevB.33.551}.

\bibitem[Stefanucci and Leeuwen(2013)]{stefanucci}
Gianluca Stefanucci and Robert~Van Leeuwen.
\newblock \emph{Nonequilibrium Many-Body Theory of Quantum Systems: A Modern Introduction}.
\newblock Cambridge University Press, 2013.

\bibitem[Sánchez et~al.(2016)Sánchez, Sothmann, and Jordan]{sanchezPE2016}
Rafael Sánchez, Björn Sothmann, and Andrew~N. Jordan.
\newblock Reprint of : Effect of incoherent scattering on three-terminal quantum hall thermoelectrics.
\newblock \emph{Physica E: Low-dimensional Systems and Nanostructures}, 82:\penalty0 359--365, 2016.
\newblock ISSN 1386-9477.
\newblock \doi{https://doi.org/10.1016/j.physe.2016.05.023}.
\newblock URL \url{https://www.sciencedirect.com/science/article/pii/S1386947716304234}.
\newblock Frontiers in quantum electronic transport - In memory of Markus Büttiker.

\bibitem[Toyabe et~al.(2010)Toyabe, Sagawa, Ueda, Muneyuki, and Sano]{Toyabi2010_Szilard_exp}
Shoichi Toyabe, Takahiro Sagawa, Masahito Ueda, Eiro Muneyuki, and Masaki Sano.
\newblock Experimental demonstration of information-to-energy conversion and validation of the generalized {{Jarzynski}} equality.
\newblock \emph{Nature Physics}, 6\penalty0 (12):\penalty0 988--992, December 2010.
\newblock \doi{10.1038/nphys1821}.

\bibitem[Whitney(2016)]{whitney2016}
Robert~S. Whitney.
\newblock \emph{Entropy}, 18:\penalty0 208, 2016.

\bibitem[Whitney et~al.(2016)Whitney, Sánchez, Haupt, and Splettstoesser]{whitneythermo1}
Robert~S. Whitney, Rafael Sánchez, Federica Haupt, and Janine Splettstoesser.
\newblock Thermoelectricity without absorbing energy from the heat sources.
\newblock \emph{Physica E: Low-dimensional Systems and Nanostructures}, 75:\penalty0 257--265, 2016.
\newblock ISSN 1386-9477.
\newblock \doi{https://doi.org/10.1016/j.physe.2015.09.025}.
\newblock URL \url{https://www.sciencedirect.com/science/article/pii/S1386947715302083}.

\end{thebibliography}

\end{document}